\documentstyle{paper}
\begin{document}

\title{2-2 AND 2-3 INTERCOMBINATION TRANSITIONS IN BE-LIKE IONS}

\author{Yuri V. RALCHENKO\\
{\it Department of Particle Physics, Weizmann Institute of Science,
Rehovot 76100, ISRAEL}\\
\\
Leonid A. VAINSHTEIN\\
{\it P.N.Lebedev Physical Institute, Moscow 117924, RUSSIA}}

\maketitle

\section*{Abstract}
We report here on calculation of probabilities of intercombination transitions
$2s^2$ $^1S_0$ - $2s2p$ $^3P_1$, $2s3p$ $^3P_1$ for Be-like ions along the
isoelectronic sequence for large range of $Z$.  Our results obtained with the
$1/Z$ expansion method agree well with experimental data including the recent
ones.

\section{Introduction}
Intercombination, or spin-forbidden, transitions are due to deviations from
the pure LS-coupling. Since their the very existence, at least for small
nuclear charges $Z$, is a consequence of the subtle relativistic spin-orbit 
interaction, the agreement between theoretical and experimental results is an 
important measure of our understanding of atomic structure and interactions. 
In addition, intercombination lines are very important for plasma diagnostics, 
for example, in astrophysics and tokamak studies [8].

Since spin-forbidden transitions for not too large $Z$ are much weaker than
spin-allowed ones, the reliable determination of their transition probabilities
requires serious experimental efforts.
Recently two important experiments on measurement of the intercombination
transitions probabilities in Be-like ions were carried out. Kwong {\it et al}
[19] have improved their ion trap experiment on the transition 
$2s^2 \ ^1S_0$ - $2s2p \ ^3P_1$ $1909$ \AA\ in C III and
obtained the value of $121\pm 7$ $s^{-1}$ which differs considerably from
the preliminary result of the same group $75$ $s^{-1}$ [26].
In another experiment Granzow {\it et al} [14] have measured the 
probability of the $2s^2 \ ^1S_0$ - $2s3p \ ^3P_1$ transition in Na VIII -- 
Si XI ions. Similar measurements for smaller nuclear charges $Z$ from N IV 
to Ne VII [9,15] were done more than ten years ago, 
so it is possible now to compare theoretical predictions along larger
interval of $Z$. These two latest experimental works initiated a wave of
theoretical calculations [11,13,3,28,12]
where most sophisticated and elaborated  up to now
methods were used, e.g. multiconfiguration Hartee-Fock (MCHF),
multiconfiguration Dirac-Fock (MCDF), configuration interaction (CI) methods,
multiconfiguration relativistic random-phase approximation (MCRRPA).
In this paper we report the results of calculations of
$2s^2 \ ^1S_0$ - $2s2p \ ^3P_1$ and $2s^2 \ ^1S_0$ - $2s3p \ ^3P_1$ transition 
probabilities along [Be] isoelectronic sequence by
the $Z$ expansion method. This approach is known to be very accurate in
determination of energy levels of few-electron multicharged ions but as we
will see in what follows gives also good results for intercombination 
probabilities.

In these calculations we use the MZ code [25] based on the perturbation 
theory on $1/Z$ parameter. The main principles of this approach and the code 
structure are well described in [25] so we refer the 
reader to this book for details. 

\section{Results and Discussion}
Both intercombination transitions studied in this paper arise from mixing of
triplet terms with the singlet ones. This mixing not only makes the
spin-forbidden transition to be possible but also influences the level energies.
Therefore the correspondence between the experimental and theoretical energies
is an additional important check for accuracy of the calculations. Earlier the
MZ method was successfully applied for calculation of the level energies for
configurations $1s^22l^{\prime }nl^{\prime \prime }$ $(n=2,3,4)$ of Na VIII - S
XIII ions [24,1]. These references contain detailed examination of energy
calculations by MZ so here we immediately pass to the discussion of obtained
results. Here we calculated energies for configurations $2l^{\prime }2l^{\prime
\prime }$ and $2l^{\prime }3l^{\prime \prime }$ for $Z=6-26$. In most cases the
difference between calculated in this work and experimental [17] energies is of
order of a few units of $10^{-4}$ or less which is typical for calculations with
the MZ code.  For intercombination transitions two energy differences, namely
$\Delta E(^3P_1-^1S_0)$ and $\Delta E(^1P_1-^3P_1)$, play a crucial role since
these quantities explicitly enter the formula for transition probability in the
first order of perturbation theory. In fact, our results for these energies have
the highest accuracy except of the latest variational calculations [29]. Note that
experimental energies for levels $2s2p$ $^3P_1$ $\left(Z=18-26\right)$ have a
stable non-depending on $Z$ shift of $\sim 350$ cm$^{-1}$ comparing to our
results. Since all these energies were determined from the same experiment [4],
it is quite possible that there was a systematic error in those measurements.

Until now only four experiments on measurement of $2s^2$ $^1S_0$ - $2s2p$ 
$^3P_1$ transition
probability were carried out, that is for C III [19], Fe XXIII
[5], Kr XXXIII [6] and Xe LI [21]. 
The best known in this list is the widely used in solar plasma
diagnostics intercombination line 1909 \AA\ in C III which has been
discussed in many theoretical papers (see references in Ref. [27]). 
As we have noted above, the recently reported experimental 
results on C III [19] do not overlap with the previous measurements of
the same group for this transition probability [26]. The new
value $A_{CIII}=121\pm 7$ $s^{-1}$ agrees well with our result $120$ $%
s^{-1}$ and MCRRPA value of 118 $s^{-1}$ [3] 
but contradicts the latest CI [11], MCHF [13] and MCDF 
[28] calculations which give $104\pm 4$, $103\pm 3$, and $100.3\pm 4$ 
$s^{-1}$ respectively. As one can see from Fig. 1, the agreement
between MZ probabilities and available data for Fe, Kr and Xe is also good. 
In addition to above mentioned CI, MCDF, MCHF, and MCRRPA calculations, we
show on this Figure the extensive MCDF calculations for Be-like ions up 
to Xe LI [2], model potential results [20] and MC 
calculations with SUPERSTRUCTURE code [22] as
well. Although different theoretical results give practically the same both 
$Z$-dependence and absolute values of probabilities for large $Z\geq26$, the 
situation for small nuclear charges is not very clear yet and new accurate 
measurements for $Z<10$ would be of great importance.

On Fig. 2 the theoretical and experimental results for transition probability
$A(2s^2$ $1S_0-2s3p$ $^3P_1)$ for $Z=7-26$ are shown. At present there are old
experimental data for $Z=7-9$ [9], $Z=10$ [15] and latest data for $Z=11-14$
[14]. The theoretical results on Fig. 2 include MCDF [12], MCHF [7], CI [16] and
HF-with-relativistic-corrections [10] calculations. As was shown by Fritzsche
and Grant [12], for small $Z$ the MCDF method is rather slow convergent with
increase of the number of configurations included. This is probably to be a
reason for large discrepancy between that paper and results of Kim {\it et al}
[18]. It is seen from Fig. 2 that different theoretical approaches show again
similar $Z$-dependence, except for smallest $Z$, where more accurate
measurements would be desirable. The new experimental results by Granzow {\it et
al} [14] seem to give $Z$-dependence rather different from theoretical one,
although in sufficiently narrow region. At least for Na VIII the value of $A$ is
considerably larger than most recent theoretical results. This feature needs
further investigation, especially since the authors claim that this value
appears to be an overestimate.

\section{Acknowledgments}
We are grateful to Drs. S.Fritzsche and I.P.Grant for making their results
available prior to publication and to Drs. P.L.Smith and E.Tr\"abert 
for interesting discussions. Yu.V.R. acknowledges the financial support
from Israeli Ministries of Absorption and Science and hospitality of 
Dr. I.Yu.Tolstikhina during his stay in Nagoya.

\vspace{1pc}
\re
 1.	Ando, K. et.al., 1992, {\it Phys.\ Scr.}, {\bf 46}, 107.

\re
 2.	Cheng, K.T. et.al., 1979, {\it At.\ Data\ Nucl.\ Data\ 
Tables} {\bf 24}, 111.

\re
 3.	Chou, H.-S. et.al., 1994, {\it Chin.\ J.\ Phys.} {\bf 32}, 261.

\re
 4.	Dere, K.P., 1978, {\it Astrophys.\ J.} {\bf 221}, 1062.

\re
 5.	Dietrich, D.D. et.al., 1978, {\it Phys.\ Rev.\ A} {\bf18}, 
208.

\re
 6.	Dietrich, D.D. et.al., 1980, {\it Phys.\ Rev.\ A} {\bf22}, 1109.

\re
 7.	Ellis, D.G., 1983, {\it Phys.\ Rev.\ A} {\bf28}, 1223.

\re
 8.	Ellis, D.G. et.al.,1990, {\it Comm.\ At.\ Mol.\ Phys.}{\bf 22}, 241.

\re
 9.	Engstr\"om, L. et.al., 1979, {\it  Phys.\ Scr.} {\bf 20}, 88.

\re
10.	Fawcett, B.C.,1985, {\it At.\ Data\ Nucl.\ Data\ Tables} {\bf 33}, 
479.

\re
11.	Fleming, J. et.al., 1994, {\it Phys.\ Scr.} {\bf 49}, 316.

\re
12.	Fritzsche, S. and Grant, I.P., 1994, {\it Phys.\ Scr.} {\bf 50}, 473.

\re
13.	Froese Fischer, C., 1994, {\it Phys.\ Scr.} {\bf 49}, 323.

\re
14.	Granzow, U. et.al.,1994, {\it  Phys.\ Scr.} {\bf 49}, 148.

\re
15.	Hardis, J.E. et.al., 1983, {\it  Phys.\ Rev.\ A} {\bf 27}, 257.

\re
16.	Hibbert, A., 1979, {\it J.\ Phys.\ B} {\bf 12}, L661.

\re
17.	Kelly, R.L., 1987, {\it J.\ Phys.\ Chem.\ Ref.\ Data}, 
{\bf 16}, Suppl. 1.

\re
18.	Kim, Y.-K. et.al., 1988, {\it J.\ Opt.\ Soc.\ Am.\ B }
{\bf 5}, 2215.

\re
19.	Kwong, V.H.S. et.al., 1993 {\it Astrophys.\ J.} {\bf 411}, 431.

\re
20.	Laughlin, C. et.al., 1978, {\it J.\ Phys.\ B} {\bf 11}, 2243.

\re
21.	M\"oller, G. et.al., 1989, {\it Z.\ Phys.\ D} {\bf 11}, 333.

\re
22.	Nussbaumer, H. and Storey, P.J., 1979, {\it J.\ Phys.\ B} 
{\bf 12}, 1647.

\re
23.	Safronova, U.I. and Senashenko, V.S., 1982, {\it Phys.\ Scr.}, {\bf 25},
 37.

\re
24.	Safronova, U.I. et.al., 1990, {\it Optics\ and\
Spectroscopy} {\bf 68}, 151.

\re
25.	Shevelko, V.P. and Vainshtein, L.A., 1993, {\it Atomic\ Physics\ 
for\ Hot\ Plasmas} \/(IOP Publishing Ltd., Bristol).

\re
26.	Smith, P.L. et.al., 1984,{\it  Phys.\ Scr.} {\bf 18}, 88.

\re
27.	Tr\"abert, E., 1993, {\it Phys.\ Scr.} {\bf 48}, 699.

\re
28.	Ynnerman, A. and Froese Fischer, C., 1995, {\it Phys.\ Rev.\ A}, 
{\bf 51}, 2020.

\re
29.	Zhu, X.-W. and Chung, K.T., 1994, {\it Phys.\ Rev.\ A} {\bf 50}, 3818.

\newpage
\begin{figure}[t]
\vspace{15pc}
\caption{Scaled transition probabilities for $2s^2 \ ^1S_0$ - $2s2p \ ^3P_1$ 
line in Be-like ions. 
{\it Experiment}\/: $\bullet$ Z=6 \ [19], 
Z=26 \ [5], Z=36 \ [6], Z=54 \ [21]; 
{\it Theory}\/: --- this paper, $\triangle$ \ [20], 
$\Diamond$ \ [22], $\cdot\cdot\cdot$ \ [2], 
+ \ [11], $\times$ \ [13], -- -- -- \ [3], 
$\Box$ \ [28].}
\end{figure}

\begin{figure}[t]
\vspace{14pc}
\caption{Scaled transition probabilities for $2s^2$ $^1S_0$ - $2s3p$ $^3P_1$
line in [Be] ions. 
{\it Experiment}\/: $\bullet$ Z=7-9 \ [9], 
Z=10 \ [15], Z=11-14 \ [14]; 
{\it Theory}\/: --- this paper, $\triangle$ \ [16], 
$\Diamond$ \ [7], -- -- -- \ [12], $\cdot\cdot\cdot$ \ [10].} 
\end{figure}

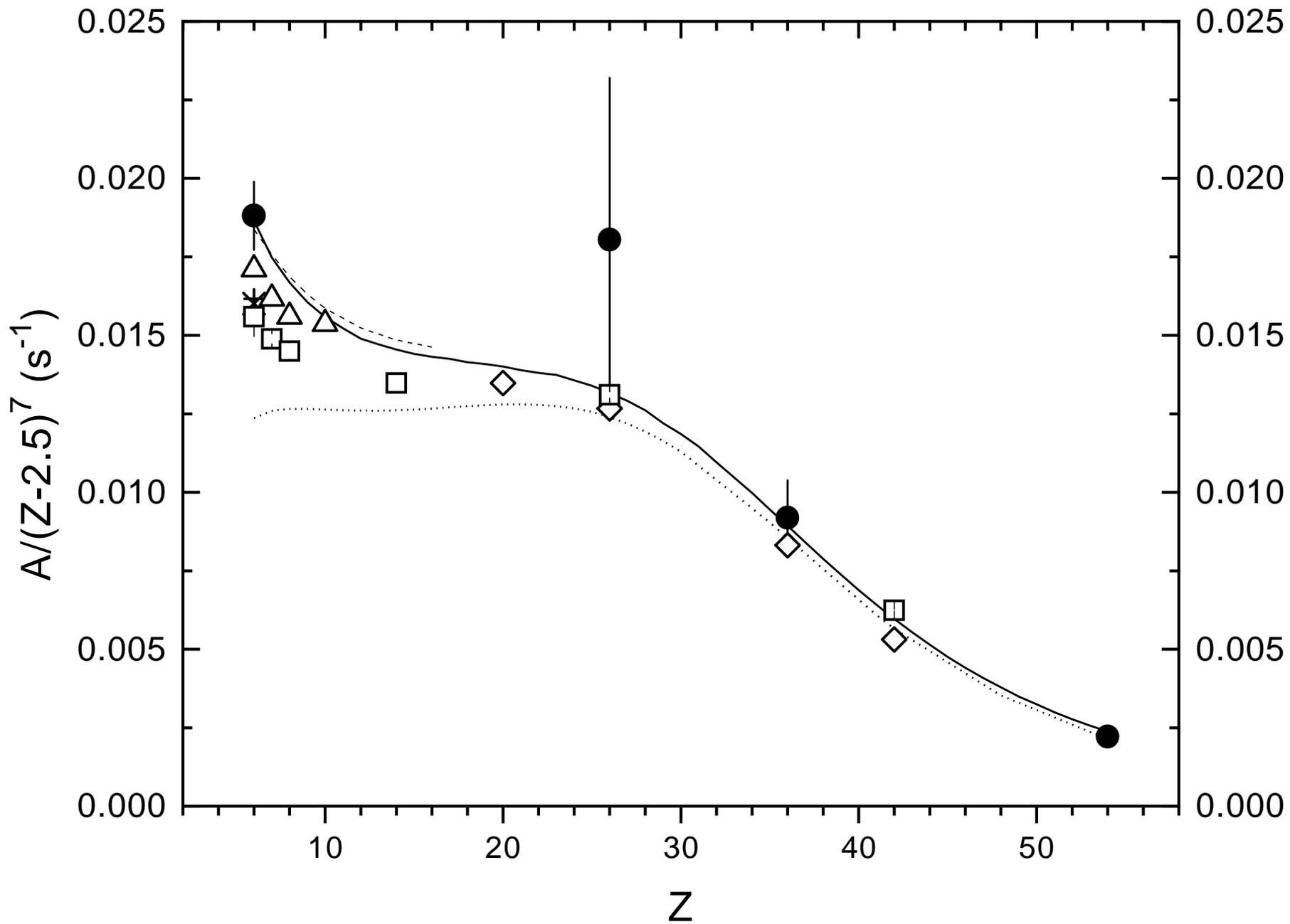
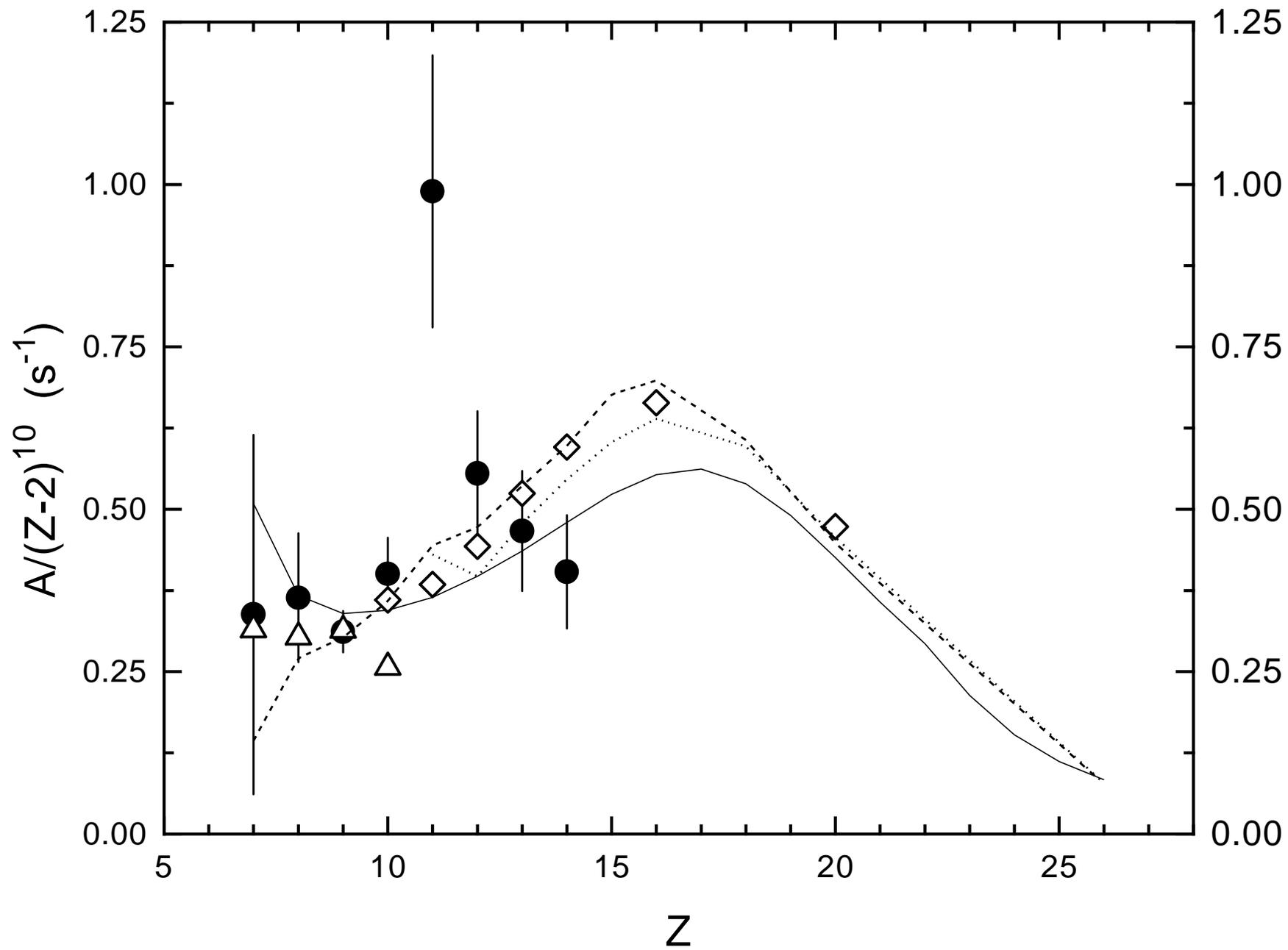
\end{document}